# Inverse Doppler Effects in Pipe Instruments

S. L. Zhai[1,3], J. Zhao[2,3], F. L. Shen[1], L. L. Li[1] & X. P. Zhao[1]

[1]Smart Materials Laboratory, Department of Applied Physics, Northwestern Polytechnical University, Xi'an 710129 P. R. China. [2]Medtronic plc, Boulder, CO 80301, USA. [3]These authors contributed equally to this work. Correspondence and requests for materials should be addressed to J.Z. (email: zhaojing1120@gmail.com) and X.P.Z. (email: xpzhao@nwpu.edu.cn).

Music is older than language, and for most of human history music holds our culture together. The pipe instrument is one of the most popular musical instruments of all time. Built on the foundation of previous flute and flute-like acoustic metamaterial models, we herein report the experimental results of the inverse Doppler effects discovered in two common pipe instruments - recorder and clarinet. Our study shows that the inverse Doppler effects can be detected at all seven pitches of an ascending musical scale when there is a relative motion between a microphone (observer) and abovementioned two pipe instruments (source). The calculated effective refractive indices of these two pipe instruments are negative and varying across a set of pitches, exhibiting a desired characteristic of broadband acoustic metamaterials. This study suggests that recorder and clarinet may be the earliest man-made acoustic metamaterials known so far, offering a new explanation why pipe instruments have enjoyed wide popularity in Europe and Asia over the past hundreds and thousands years. This newly discovered phenomenon would also offer a clue into designing next-generation smart broadband double-negative acoustic metamaterials with varying refractive index.

The Doppler effect is a fundamental phenomenon in wave propagation. In 1843, Doppler first generalized the change in frequency of a wave for an observer moving relative to its source[1,2]. Nowadays, this phenomenon is utilized in many fields, including space technology, traffic control, disease diagnosis, to name a few[3-6]. In 1968, Veselago theoretically predicted that the inverse Doppler effect[7] could be observed in materials with negative refractions ($n<0$). In the 1990s, Pendry *et al.*

presented that periodically arranged metallic wires and split resonance rings can realize negative permittivity and negative magnetic permeability, respectively[8, 9]. In 2001, Smith *et al.* first manufactured a left-handed material with periodically arranged metallic wires and split rings, and experimentally demonstrated the phenomenon of negative refraction[10]. Since then the research of left-handed materials boomed, resulting in electromagnetic metamaterials[11-15] being one of the mostly studied area. In 2000, Liu *et al.* proposed the idea of localized resonance, which later became the foundation of acoustic metamaterials. In particular, each resonance unit - which is equivalently an artificial "meta-atom" - that constitutes the acoustic metamaterial is orders of magnitude smaller in size relative to the effective operating wavelength. The resonant frequency of each unit is primarily determined by its geometry and size. Various acoustic metamaterials with negative effective mass density were engineered by periodically arranging the acoustic "meta-atoms"[16]. In 2006, Fang *et al.* manufactured a one-dimensional acoustic metamaterial with negative effective dynamic modulus at ultrasound frequencies. This type of metamaterial is composed of Helmholtz resonators of which the group and phase velocities are opposite in direction near the resonant frequency. It was further showed that the dynamic bulk modulus of this metamaterial calculated based on a homogeneous medium model is negative at the resonant frequency[17]. In our previous works, we designed two types of artificial "meta-atoms" (i.e. split hollow sphere (SHS)[18, 19] and hollow steel tube (HST)[20, 21]) to achieve negative effective bulk modulus and negative mass density, respectively. Later, a double-negative acoustic metamaterial was engineered by combining these two "meta-atoms"[20, 22]. We also designed a flute-like "meta-molecule" (i.e. perforated hollow tube (PHT)[22, 23]) by seamlessly integrating two types of "meta-atoms" to achieve simultaneous negative bulk modulus and negative mass density. We demonstrated slab focusing and negative refraction effects using this double-negative metamaterial. Furthermore, based on the mechanism of weak interaction among "meta-molecules" and "meta-atoms", our team presented the concept of "meta-cluster" acoustic metamaterial[21, 24]. The mass density and modulus of the fabricated metamaterial are both negative in all directions over a broad band of frequencies. More importantly, the inverse Doppler shifts were detected at all frequencies, matching well with the calculated results derived from effective refractive index. We found that the Doppler effect in this medium altered from regular to inverse then back to regular throughout the relative motion between the observer and source[24].

The pipe instrument is one of the most popular musical instruments of all time. Flutes, as one of the earliest known instruments, dated back to 8000 years ago[25], or perhaps earlier[26], thanks to its simple construction, light weight and euphonious tones. Although our ancestors were unaware of the operating principles of a resonator, they did notice that sound pitches produced by a flute were somehow related to the arrangement of holes as well as the instrument length. The materials used to make flutes have evolved from ancient animal femurs to bamboo/metal in modern days, and their manufacturing process has become fairly sophisticated. It is now well-known that the tube length and opening manner of a flute determine its resonant frequency.

However, little attention has been paid to the sound effect caused by the relative motion between a flute and the audience.

Previously, using a standard acoustic test method, we observed the inverse Doppler effect associated with a flute at a variety of pitches[24]. Specifically, the received frequency detected by the observer (a moving microphone) decreases as it approaches the stationary source (a flute). Conversely, when the observer moving away from the source, the detected acoustic signal shifts towards higher frequencies. At an appropriate speed, the inverse Doppler effects were observed at all seven pitches (corresponding to an ascending musical scale) produced by either the blow hole or finger holes of the flute, in spite of their differences in sound intensity and frequency. Higher pitch is associated with a greater shift in frequency. The calculated refractive indices at these pitches are all negative (within the range from -0.6 to -2)[24] when the inverse Doppler effects occurred. In this paper, the inverse Doppler effects of sound pitches produced by the blow holes of a recorder and a clarinet are presented. The aim is to expand our previous conclusion about the attractive metamaterial properties of a flute into other popular pipe instruments. This generalization would provide a new direction for designing the basic units of next generation smart broadband acoustic metamaterials.

**Results**

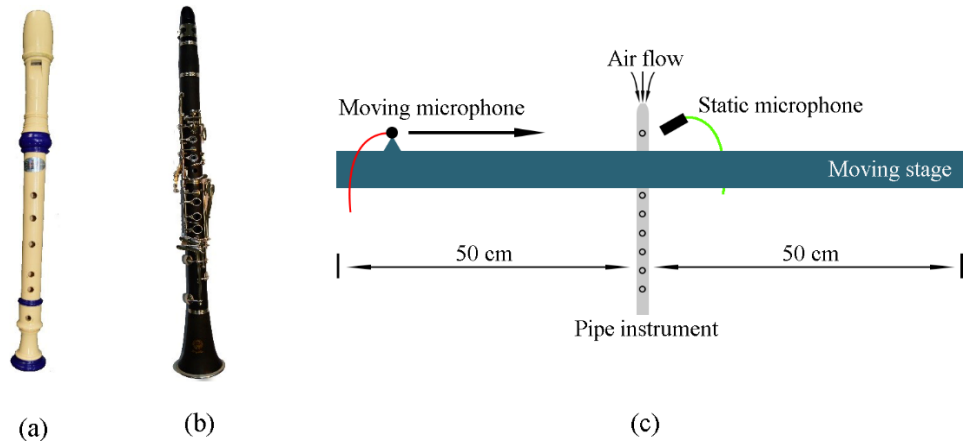

**Figure 1 a** and **b** Photographs of a recorder and a clarinet, respectively. **c**, Experimental setup of the Doppler shift.

Figures 1a and 1b show the photographs of the recorder and clarinet used in our experiments, respectively. A schematic diagram of the Doppler shift experiment is shown in Fig. 1c. We used a flexible tube connected with an electric air pump to blow steady air stream into the blow hole to ensure continuous acoustic waves coming out from the pipe instrument. For the sake of simplicity, the pipe instrument is secured at a fixed location. A microphone is placed next to its blow hole to detect the sound signal produced by this stationary source. The fundamental frequency of this signal is used as the baseline in our frequency shift analysis. A second microphone is mounted on a 1-D motorized translation stage moving towards and away from the source at a constant speed. This microphone serves as the observer in this experiment. Note that abovementioned two microphones receive the acoustic signal produced by the instrument at the same time. Based on the waveform captured by the stationary microphone, through the LabVIEW Signal Express Software, we can then calculate the frequency of the source at a given pitch, namely the baseline frequency. Similarly, using the waveform detected by the moving microphone, we can obtain the shifted frequency when there is a relative motion between the source and observer. The data processing of the Doppler shift for the pipe instruments is the same as that for the metamaterial samples presented in[24].

Figure 2 visualizes the sound signal (pitches 2, 4, and 6) from a recorder detected by the moving microphone at a speed of 0.5 m/s and 0.1 m/s. The black curve is the waveform detected by the stationary microphone, while the red curve refers to the waveform captured by the moving microphone. The static frequency and moving frequencies (including approaching and departing) are also labeled in Fig 2. The total moving distance of the microphone in our experiment is 100 cm, because the length of the motorized translation stage is 100 cm. Massive data processing demonstrated that the results of frequency shift are identical during the whole approaching region, so is that during the receding region. Figure 2 only shows partial fragments of both approaching and receding process to calculate the frequency shifts. It is found for all of the pitches that, the received signal shifts towards lower frequencies when the observer approaching the source. As the observer moving away from the source, the received frequency increases. Hence, the inverse Doppler effect is observed in the sound waves produced by the recorder. Besides, the faster the moving speed is, the larger the frequency shift is.

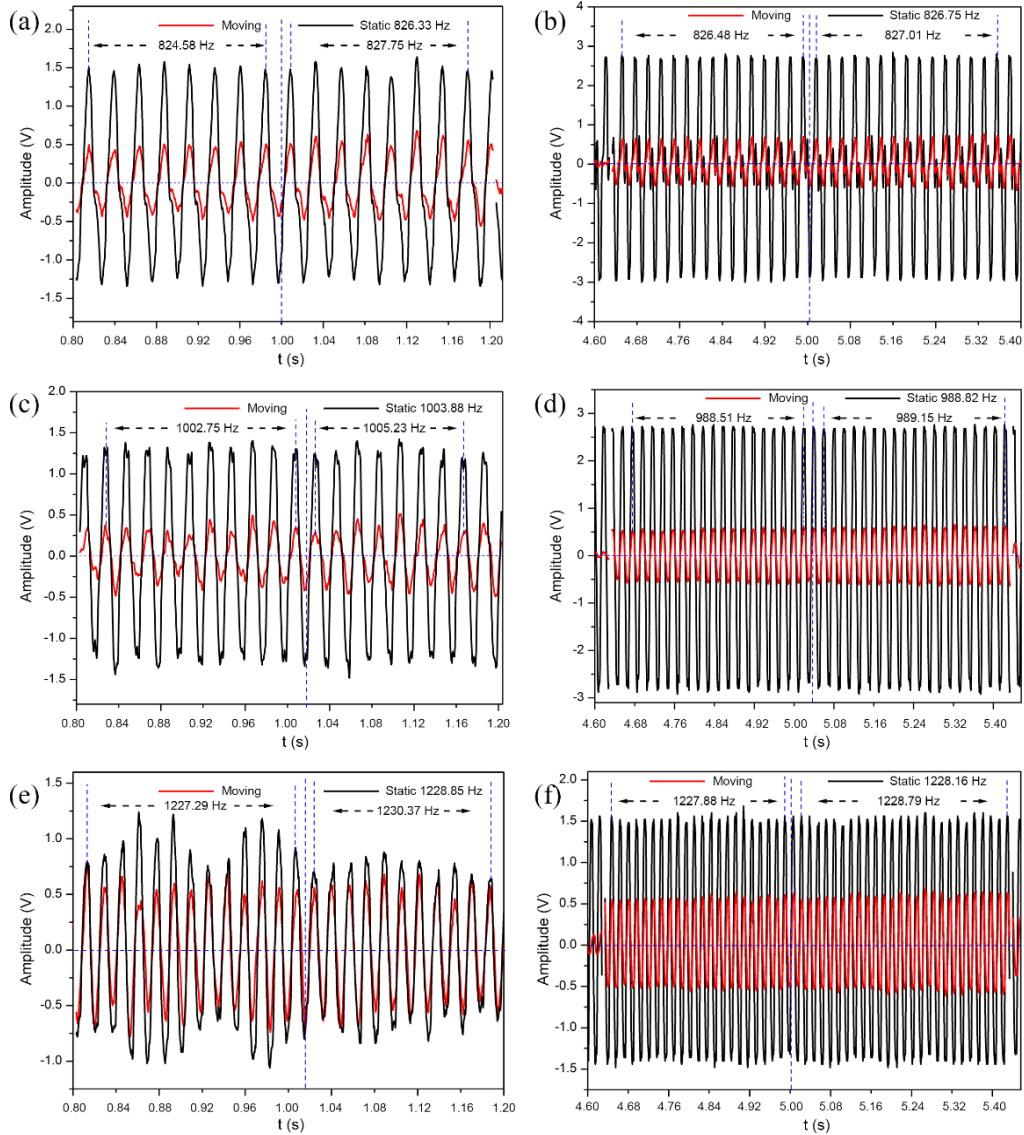

**Figure 2 Measured Doppler shift results of the recorder for different pitches at different moving speeds. a** and **b** are the results for pitch 2 at the speed of 0.5 m/s and 0.1 m/s, respectively. When the moving microphone approaches the source, the detected frequencies are reduced by 1.75 Hz and 0.27 Hz, respectively; as the microphone recedes, the frequencies increase by 1.42 Hz and 0.26 Hz, respectively. **c** and **d** are the results for pitch 4 at the speed of 0.5 m/s and 0.1 m/s, respectively. When the moving microphone approaches the source, the detected frequencies are reduced by 1.13 Hz and 0.31 Hz, respectively; as the microphone recedes, the frequencies increase by 1.35 Hz and 0.33 Hz, respectively. **e** and **f** are the results for pitch 6 at the speed of 0.5 m/s and 0.1 m/s, respectively. When the moving microphone approaches the source, the detected frequencies are reduced by 1.56 Hz and 0.28 Hz, respectively; as the microphone recedes, the frequencies increase by 1.52 Hz and 0.63 Hz, respectively.

We systematically studied the Doppler shift of the recorder in terms of pitch

(frequency) and phonation position. To further analyze the Doppler behavior pertain to a recorder, Figure 3 presents the measured baseline frequency of each individual pitch as well as the frequency shift $\Delta f$. The refractive index is calculated according to the formula $n= \pm \Delta f \cdot v/(f \cdot v_{observer})$. Take pitch 4 for example, the measured baseline frequency is 1231.16 Hz. This frequency is lowered to 1229.90 Hz when the observer moving towards the source. Once passed, it shifts higher to 1232.24 Hz. Thus, the frequency shift in approaching is -1.26 Hz, and the corresponding refractive index is -0.70. The frequency shift in departing is 1.08 Hz, and the corresponding refractive index is -0.60. In theory, the absolute value of frequency shift and refractive index are independent of the direction of motion at a given speed, as is verified in Fig. 3 accounting for measurement uncertainties. It is also worth noting that the seven pitches tested here forms a complete ascending scale. The inverse Doppler effects are observed over this entire broad band of frequencies. So as the negative refractive indices, as shown in Table 1. Therefore, the recorder itself is a type of broadband acoustic metamaterial.

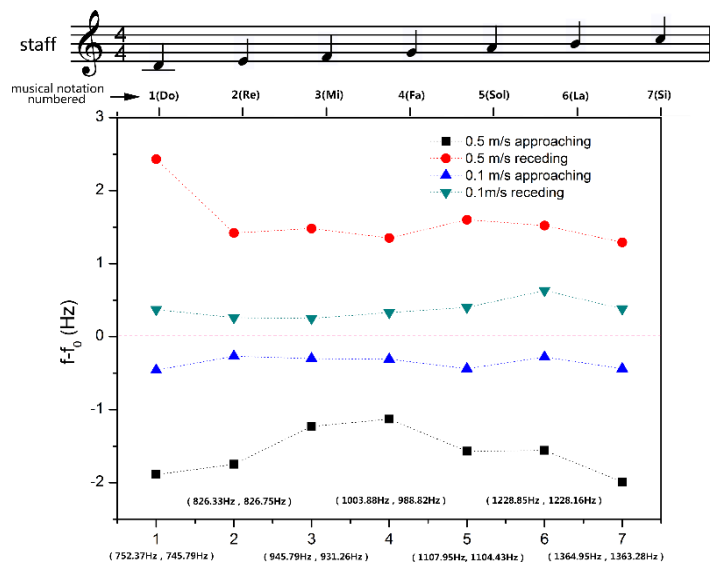

**Figure 3 Measured Doppler shift results of the recorder for individual pitches (i.e. 1-7).** The black square and red circle refer to the frequency shift results of approaching and receding processes, respectively, which are measured at the speed of 0.5 m/s. The blue triangle and cyan

inverted triangle indicate the frequency shift results of approaching and receding processes, respectively, which are measured at the speed of 0.1 m/s. The static frequencies of the recorder at the speeds of 0.5 m/s and 0.1 m/s for individual pitches are labeled in the brackets. The positive values mean the detected frequency is larger than source frequency, and the negative values imply the contrary situation.

Figure 4 displays the sound signal (pitches 2, 3, and 5) detected by the moving microphone at a speed of 0.5 m/s and 0.1 m/s with a clarinet as the device under test. The waveform detected by the stationary microphone is shown in black, while the red curve refers to the waveform captured by the moving microphone. Figure 4 only shows partial fragments of both approaching and receding process to calculate the frequency shifts. As seen, received signal shifts towards lower frequencies when the observer approaching the source. In contrast, as the observer departing from the source, the received frequency increases. Hence, the clarinet also exhibits the interesting characteristic of the inverse Doppler effect.

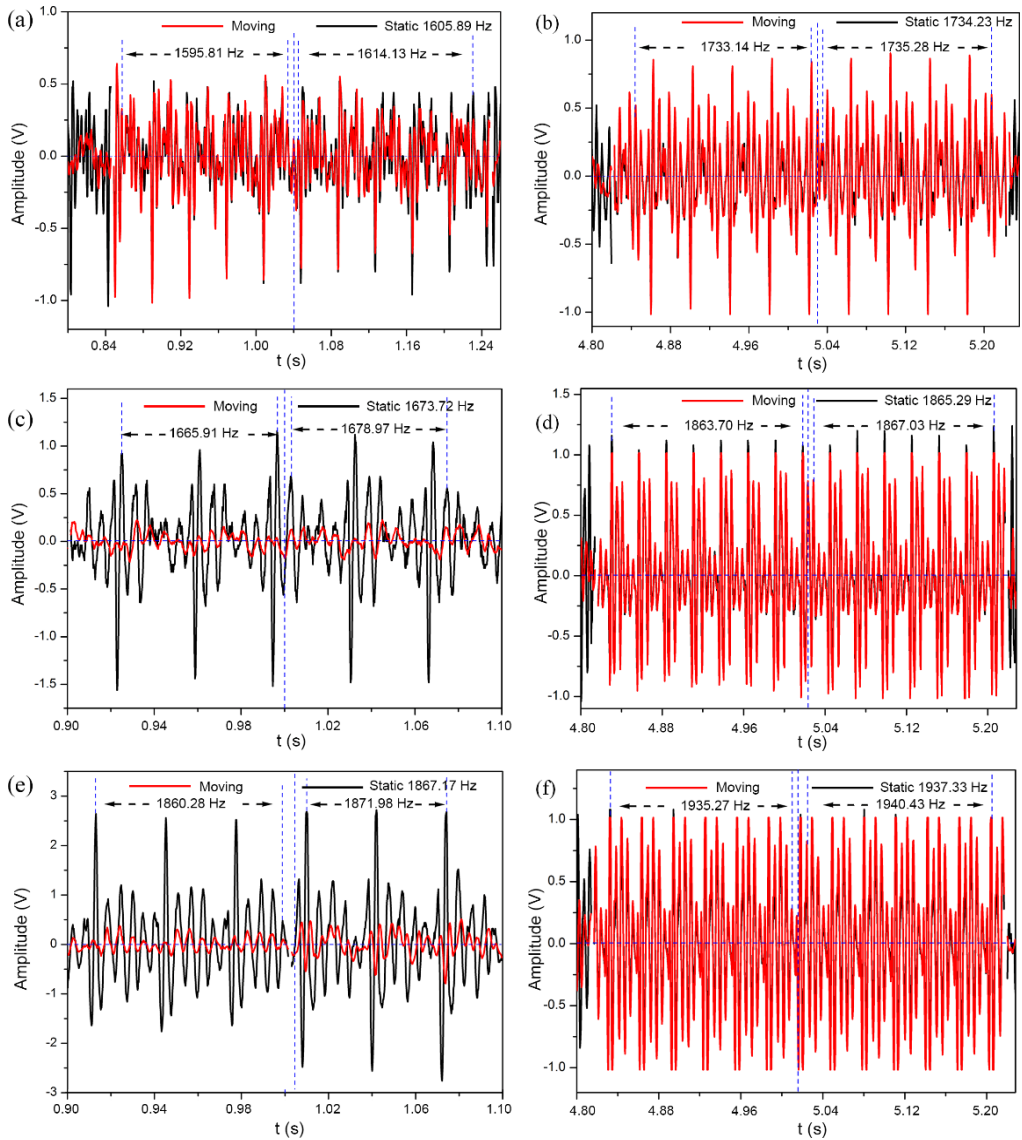

**Figure 4 Measured Doppler results of the clarinet for different pitches at different moving speeds. a** and **b** are the results for pitch 2 at the speed of 0.5 m/s and 0.1 m/s, respectively. When the moving microphone approaches the source, the detected frequencies are reduced by 10.08 Hz and 1.09 Hz, respectively; as the microphone recedes, the frequencies increase by 8.24 Hz and 1.05 Hz, respectively. **c** and **d** are the results for pitch 3 at the speed of 0.5 m/s and 0.1 m/s, respectively. When the moving microphone approaches the source, the detected frequencies are reduced by 7.81 Hz and 1.59 Hz, respectively; as the microphone recedes, the frequencies increase by 5.28 Hz and 1.74 Hz, respectively. **e** and **f** are the results for pitch 5 at the speed of 0.5 m/s and 0.1 m/s, respectively. When the moving microphone approaches the source, the detected frequencies are reduced by 6.89 Hz and 2.06 Hz, respectively; as the microphone recedes, the frequencies increase by 4.81 Hz and 3.1 Hz, respectively.

Figure 5 shows the Doppler effect test results of the clarinet at different pitches and speeds. The inverse Doppler effects are observed across all seven frequencies of a musical scale. The calculated refractive indices of clarinet are all negative over the same range of frequencies, as shown in Table 1. Therefore, the clarinet is also a type of acoustic metamaterial. From Fig. 3 and Fig. 5, insignificant differences in refractive index are seen between record and clarinet, but the frequency shifts of clarinet appear to be greater than that of recorder.

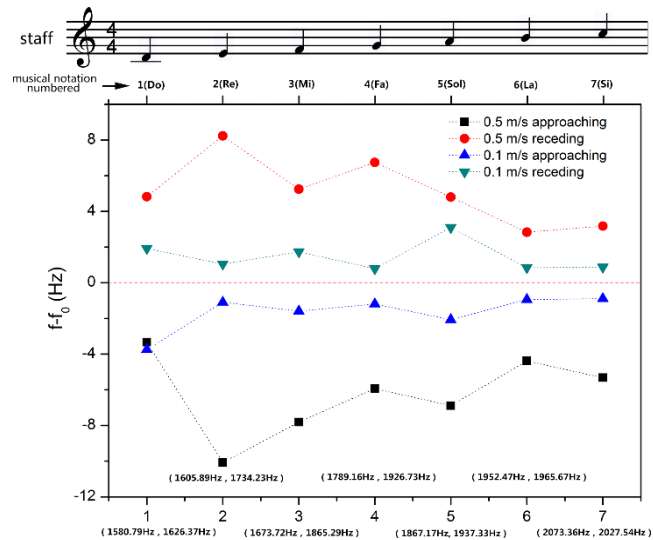

**Figure 5 Measured Doppler shift results of the clarinet for individual pitches (i.e. 1-7).** The black square and red circle refer to the frequency shift results of approaching and receding processes, respectively, which are measured at the speed of 0.5 m/s. The blue triangle and cyan inverted triangle indicate the frequency shift results of approaching and receding processes, respectively, which are measured at the speed of 0.1 m/s. The static frequencies of the clarinet at the speeds of 0.5 m/s and 0.1 m/s for individual pitches are labeled in the brackets. The positive values mean the detected frequency is larger than source frequency, and the negative values imply the contrary situation.

**Table 1. Refractive indexes ($n$) of the recorder and clarinet for individual tones.**

| tone | 0.5 m/s | | | | 0.1 m/s | | | |
|---|---|---|---|---|---|---|---|---|
| | $n$ (approaching) | | $n$ (receding) | | $n$ (approaching) | | $n$ (receding) | |
| | recorder | clarinet | recorder | clarinet | recorder | clarinet | recorder | clarinet |
| 1 | -1.7 | -1.45 | -1.6 | -2.10 | -2.12 | -7.91 | -1.7 | -4.05 |
| 2 | -1.5 | -4.31 | -1.2 | -3.52 | -1.12 | -2.16 | -1.08 | -2.08 |
| 3 | -0.89 | -3.20 | -0.65 | -2.15 | -1.1 | -2.92 | -0.92 | -3.20 |
| 4 | -0.77 | -2.27 | -0.53 | -2.59 | -1.08 | -2.12 | -1.14 | -1.41 |

| 5 | -0.97 | -2.53 | -0.6 | -1.76 | -1.37 | -3.65 | -1.24 | -5.49 |
| 6 | -0.87 | -1.54 | -0.49 | -1.00 | -0.78 | -1.64 | -1.76 | -1.48 |
| 7 | -1 | -1.76 | -0.5 | -1.05 | -1.11 | -1.51 | -0.96 | -1.49 |

Thanks to its locally resonant structure, flute-like artificial "meta-molecule" with negative modulus and mass density exhibits irregular but interesting properties, such as negative refraction and the inverse Doppler effect[22, 23]. The experimental results demonstrated that the inverse Doppler effects and negative refraction can be detected in a flute – a musical instrument has enjoyed popularity for centuries in Asia and Europe[24]. In addition to flute, recorder and clarinet which we have studied thus far, several other pipe and wind instruments including Xun, Sheng (as shown in Fig. 6), suona horn and oboe could also be shown to feature such unique characteristics like the inverse Doppler effect and negative refraction.

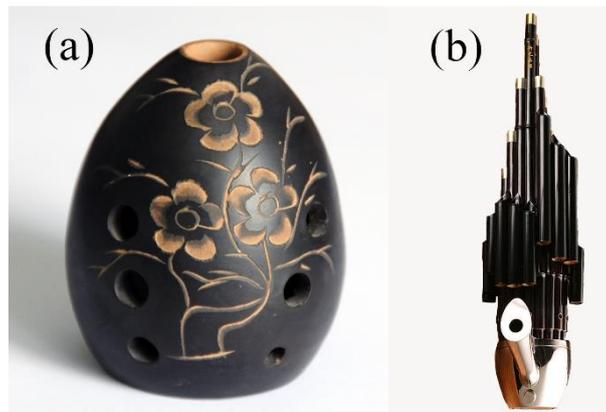

**Figure 6 Photographs of two Chinese traditional pipe instruments. a**, Xun, one of the oldest musical instruments in China and has been in use for approximately seven thousand years. **b**, Sheng, one of the oldest musical instruments in China and has been in use for approximately three thousand years.

Our observations suggest that, in the process of playing music, the motion states of instrument can be divided into three categories, namely the state of stationarity, medium-speed movement, and high-speed movement. The stationarity state will not give rise to the phenomenon of Doppler effect. The speeds of 0.1 m/s and 0.5 m/s just correspond to the states of medium-speed movement, and high-speed movement, respectively. Therefore, we chose 0.1 m/s and 0.5 m/s as moving speeds in our tests, so that the experimental results can cover major motion states of the instrument in the process of playing music.

In a typical stage performance, the location of a pipe instrument player relative to the audience is constant. However, relative motions can occur between the blow/finger holes (sound source) and the ears (receiver) of the audience. These body movements may include reciprocating motions along the horizontal, vertical, front and back, and even curvilinear directions. During the experiments, the pipe instruments are still and the microphone moves. On the contrary, during the musicians playing the recorder and clarinet, the pipe instruments move and the audience is still. But the inverse Doppler effect is identical between these two situations. As for the audience at more than 2m away, the received sound intensities may be different, but the frequencies are all the same. It has been widely believed that an instrument player's body movements are intended to express emotion and attract audience's attention. This study suggests the musician also consider the frequency shift phenomenon thanks to the inverse Doppler effect which would slightly change the beautiful sounds received through the ears of the audience.

The pipe instrument is possibly the earliest man-made acoustic metamaterials device. It was reported that a silver flute excavated in Iraq was made 4,500 years ago [27]. A bone flute unearthed in Jiahu, China was 8,000 years old[25]. A fossil of flute discovered in Germany dated back to 25,000 years ago[26]. In contrast, it wasn't until 1842 when the Doppler effect was first introduced to public, while the property of inverse Doppler effect may have been utilized for thousands of years. The bandwidth of conventional negative refraction metamaterials is inherently limited by the local resonance. However, the pipe instruments exhibit different negative refractive indices over a wide range of frequencies. It therefore can be used as a basic unit for designing various novel smart broadband acoustic metamaterials.

**Conclusion**

Inspired by the recent findings in flute and flute-like acoustic metamaterials, we systematically studied the inverse Doppler effects of the recorder and clarinet at seven sound pitches of an ascending musical scale. The experimental results show that, the detected acoustic signals produced by these two pipe instruments shift towards lower frequencies when the observer approaching the source, whereas higher frequencies are received as the observer departing from the source. The magnitude of shift in frequency varies with respect to the baseline frequency produced by the stationary instrument. The inverse Doppler effects are observed across the frequency range of an ascending musical scale (~500Hz) at two different speeds of motion. The calculated effective refractive indices of these two pipe instruments remain negative throughout the experiment at all frequencies and speeds. Therefore, these two instruments are essentially broadband acoustic double-negative metamaterials with varying index of refraction. Our study also suggests that pipe instrument players worldwide unknowingly take advantage of the inverse Doppler effect during stage performance. Their impulsive body movements slightly change the sound pitches received through the ears of the audience – which in turn offers an interesting explanation why pipe instruments have enjoyed such a wide popularity for a long time.

# Methods

**Experimental facilities.** An air pump, which can provide a stable air speed, is connected to the blow hole of the pipe instrument to generate sound signals. Two free field microphones (MPA416, BSWA, Beijing, China) are connected to an oscilloscope (Tektronix DPO3014) to receive and record the sound signals. An 1D motorized translation stage is used to provide the needed moving speed.

**Measurement method of the Doppler shifts of the pipe instruments.** The devices to generate acoustic waves are the pipe instruments themselves. For the convenience of measurement, the position of the pipe instrument is fixed. A microphone is mounted on the 1D motorized translation stage to move toward and away from the pipe instrument with a fixed speed. As the pipe instrument is driven by the air pump, the exact stationary frequencies of the sound wave generated by the pipe instrument in different tones are unknown in advance. In order to compare the stationary frequency with the moving frequency at the same time, the second microphone is fastened near the blow hole of the pipe instrument to detect the stationary signal of sound wave at the same time as the first microphone moves toward and away from the pipe instrument. The actual frequency of the pipe instrument can be obtained by calculating the wave number per unit time. [24]

## Acknowledgements

This work was supported by the National Natural Science Foundation of China under Grant Nos.11674267 and 51272215 and the National Key Scientific Program of China (under project No. 2012CB921503)

## Author contributions

X.P.Z. conceived the idea and designed the experiments; L. L. L. and F. L. S. performed the experiments; S.L.Z. and J.Z. wrote the manuscript; S.L.Z. drafted the text and aggregated the figures; X.P.Z. and J.Z. discussed the results and revised the manuscript.

## Additional information

**Supplementary information accompanies this paper at http://www.nature.com/srep**

**Competing financial interests:** The authors declare no competing financial interests.